\documentclass{aa}  

\usepackage{graphicx}
\usepackage{natbib}
\usepackage{txfonts}
\usepackage{lipsum}
\usepackage{subcaption}         
\usepackage{lscape}             
\usepackage{placeins}           
\usepackage[export]{adjustbox}
\usepackage{hyperref}
\hypersetup{
    colorlinks=true,
    linkcolor=blue,
    citecolor=blue
    }

\begin{document}

   \title{An extensive grid of DARWIN models for M-type AGB stars}

   \subtitle{II. Effects of pulsation periods on wind properties}

   \author{E. Siderud\inst{1} \corrauth{emelie.siderud@physics.uu.se}
        \and K. Eriksson\inst{1} \email{kjell.eriksson@physics.uu.se}
        \and S. Höfner\inst{1} \email{susanne.hoefner@physics.uu.se}
        \and A. Ahmad\inst{1}$^,$\inst{2}$^,$\inst{3} \email{ariefahmad@g.ecc.u-tokyo.ac.jp}
        }

   \institute{Theoretical Astrophysics, Division for Astronomy and Space Physics, Department of Physics and Astronomy, Uppsala University, Box 524, 751 20 Uppsala, Sweden
            \and Department of Astronomy, Graduate School of Science, The University of Tokyo, Tokyo 113-0033, Japan
            \and National Astronomical Observatory of Japan, Mitaka, Tokyo 181-8588, Japan
            }

   \date{Received 15 May 2026 / Accepted 29 June 2026}

  \abstract 
   {Mass loss from asymptotic giant branch (AGB) stars is the result of a complex interplay between pulsation, atmospheric dynamics, dust formation, and radiative acceleration. Pulsation periods are a key input in dynamical atmosphere and wind models, and different prescriptions for assigning periods based on stellar parameters may lead to systematic differences in the predicted wind properties.}
   {To better constrain this critical parameter, we investigated how the choice of pulsation period affects the wind properties of dynamical atmosphere and wind models of M-type AGB stars by comparing models based on an empirical period--luminosity (P--L) relation with corresponding ones that adopt a period--mean density relation derived from 3D pulsation models.} 
   {We analysed two grids of DARWIN models that cover a range of current stellar masses, luminosities, and effective temperatures. For each grid, pulsation periods were assigned using either the P--L relation or the period--mean density relation, allowing for a direct comparison of the resulting dynamical structures and wind properties for pairs of models differing by period only. }
   {Independent of the adopted period prescription, the time-averaged wind properties correlate strongly with $L_\star/M_\star$. The pulsation period affects the atmospheric dynamics through changes in the relative timing of shock propagation and dust formation, which affect both wind formation and the resulting wind properties. Shorter periods favour the onset of a wind, and models differing only in pulsation period can exhibit significantly different wind properties. }
   {The period--mean density relation provides a physically motivated alternative to the empirical P--L relation by accounting for stellar parameters beyond luminosity, and enables a more direct comparison between DARWIN models and observed Mira variables. }


   \maketitle
\nolinenumbers

\section{Introduction}

The atmospheres of cool asymptotic giant branch (AGB) stars are inherently dynamical due to large-amplitude radial pulsations (e.g.~\citealt{wood2015,trabucchi2019}), which results in a periodically expanding and contracting photosphere. The radial motions of the gas layers generate sound waves that steepen into strong shocks as they propagate outwards through the atmosphere. In the wake of these shocks, gas is lifted to regions with lower temperatures, creating favourable conditions for dust formation. The newly formed dust grains interact with the stellar radiation field, gaining outward momentum via the absorption or scattering of photons. Through collisions between dust and gas, momentum is transferred to the surrounding material, thereby driving a stellar wind with a substantial mass-loss rate that affects the observable properties and further evolution of the star (see e.g. reviews by \citealt{hofner2018}, \citealt{matthews2024}).

For M-type AGB stars, the most likely wind-driving dust species are silicates, which consist of relatively abundant chemical elements and give rise to characteristic mid-IR features observed at 10 and 18 microns (e.g. \citealt{gillett1968,woolf1969}). 
Silicate condensation starts when temperatures have fallen below the thermal stability limit of the condensate, typically at around 2 stellar radii.
The closest distance where dust can exist varies with pulsation phase, since overall atmospheric temperatures depend on the stellar luminosity. The temperatures are lowest around minimum light, favouring the formation of new dust layers close to the star in these phases. While temperature sets a threshold, the efficiency of dust grain growth depends on the gas density, which affects the collision rates of atoms and molecules with the grains. A shock wave that propagates outwards beyond the condensation distance and compresses gas into a dense wake can locally boost dust condensation rates. Therefore, the timing of shock propagation relative to the variation in the luminosity during the pulsation cycle is critical, as illustrated by detailed dynamical models (for an in-depth discussion, see e.g. \citealt{liljegren2017a}). 

This paper is the second in a series presenting an extensive grid of dynamic atmosphere and dust-driven wind models for M-type AGB stars that span a wide range of stellar parameters and predict mass-loss rates for use in evolution models. Paper~I \citep{bladh2019b} focused on the overall properties of the model grid. The aim of the present study was to investigate how different pulsation period prescriptions affect the resulting wind properties. The radiation-hydrodynamic (RHD) models were computed with the DARWIN code \citep{hofner2016}, where the effects of pulsation are simulated by variable conditions at the inner boundary, just below the photosphere. 
As discussed in more detail below (see Sect.~\ref{sect:method}), the position of the innermost mass shell varies sinusoidally during a pulsation cycle (parametrised in terms of pulsation period and velocity amplitude), accompanied by a periodical variation in luminosity with a prescribed amplitude.
We compared models that adopt an empirical period--luminosity (P--L) relation with models where the pulsation periods are determined from a period--mean density relation derived from global 3D RHD simulations of AGB stars. The resulting mass-loss rates, wind velocities, and dust formation efficiencies are analysed and discussed.

In dynamical atmosphere and wind models found in the literature (including those in Paper~I), the pulsation period is often set according to empirical P--L relations based on observations of long-period variables (e.g. \citealt{feast1989,whitelock2009a}). While these relations are observationally motivated and straightforward to apply, they do not explicitly account for dependences on stellar parameters other than luminosity. Therefore, the question arises as to what extent such relations actually give a realistic picture of pulsation effects on the atmospheres and winds of AGB stars.

As a physically motivated alternative to empirical P--L relations, \citet{ahmad2023} derived a period--mean density relation based on global 3D RHD simulations of pulsating AGB stars. The modal analysis of radial fundamental and overtone modes by \citet{ahmad2025}, together with the non-linear radial pulsation models of \citet{trabucchi2021}, supports the interpretation that pulsation periods are primarily governed by the stellar mean density.

Here, we present a new grid of DARWIN models that are based on a prescription of pulsation periods derived from 3D pulsation models but identical to the models in Paper~I in all other aspects. As mentioned above, the relative timing of shock propagation and luminosity variability affects dust formation and wind acceleration. We can therefore expect that differences in pulsation periods, for models with otherwise equal stellar parameters, will affect the wind properties. 

The paper is structured as follows: In Sect.~\ref{sect:method} we briefly describe the methods and input parameters used to produce the DARWIN models, focusing on the pulsation periods. In Sect.~\ref{sect:result} we analyse the differences between models based on the two prescriptions for pulsation periods, and in Sect.~\ref{sect:disc} we compare both types of models to observations. The conclusions are presented in Sect.~\ref{sect:concl}.

\section{Methods and model parameters} 
\label{sect:method}

As stated above, the purpose of this study was to isolate the effects of the chosen prescription of pulsation periods on the wind properties resulting from our simulations. The new models presented here are similar to the corresponding ones in Paper~I in all other aspects. Therefore, we only give a very brief overview of the modelling methods (referring to the previous paper for details) and focus on the different prescriptions of pulsation periods in the following.

\subsection{DARWIN models}

As in Paper~I, we used the 1D RHD code DARWIN \citep{hofner2016} to model the atmospheres and winds of M-type AGB stars. The code solves the equations describing the conservation of mass, momentum and energy, together with non-equilibrium dust formation and frequency-dependent radiative transfer. The output consists of snapshots of the radial structures, with mass-loss rates, wind velocities, and dust properties as direct results of the simulations. In order to keep the new models comparable to the previous grid, we used the same microphysical data for the gas and the silicate dust as described in Paper I. For a more detailed description of the DARWIN code, see \citet{hofner2016}, and references therein. 

The inner boundary of the models was placed just below the stellar photosphere (above the pulsation driving region). To simulate the 
effects of pulsation on the atmosphere and wind, a parameterised description of the expansion and contraction of the star was used
(introduced by \citealt{bowen1988a}). The variation of the position of the innermost mass shell, $R_\mathrm{in}(t)$, is given by

\begin{equation}
\label{eq:R_in}
    R_\mathrm{in}(t) = R_\mathrm{0} + \frac{\Delta u_\mathrm{p} P}{2\pi} \sin{\left(\frac{2\pi}{P}t\right)}
,\end{equation}

\noindent where $R_\mathrm{0}$ is the average radial position of the boundary, $\Delta u_\mathrm{p}$ is the velocity amplitude, and $P$ is the pulsation period. 
This corresponds to a local gas velocity of 

\begin{equation}
\label{eq:u_in}
    u_\mathrm{in}(t) = \Delta u_\mathrm{p} \cos{\left(\frac{2\pi}{P}t\right)}.
\end{equation}

\noindent The accompanying variable luminosity is given by 

\begin{equation}
\label{eq:L_in}
    L_\mathrm{in}(t) = f_\mathrm{L}\left(\frac{R_\mathrm{in}^2(t)-R_\mathrm{0}^2}{R_\mathrm{0}^2}\right)\times L_\mathrm{0} + L_\mathrm{0}
\end{equation}

\noindent (introduced in \citealt{gautschy-loidl2004}), where $L_0 =L_\star$ is the luminosity of the hydrostatic initial model (see below), and the parameter $f_\mathrm{L}$ sets the amplitude. The value of $f_\mathrm{L}$ was chosen to reproduce typical observed photometric variations.

   \begin{figure}[t!]
   \centering
   \includegraphics[width=\hsize]{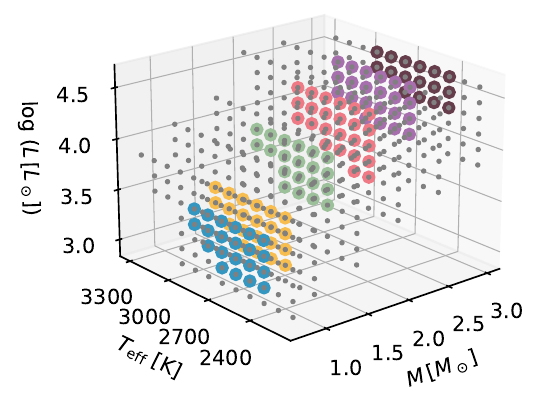}
      \caption{Overview of the stellar parameters covered in the grid in Paper~I (grey dots) compared to the sub-grid analysed in this paper (circles colour-coded by mass). }
         \label{fig:grid}
   \end{figure}

   \begin{figure}[t!]
   \centering
   \includegraphics[width=0.9\hsize]{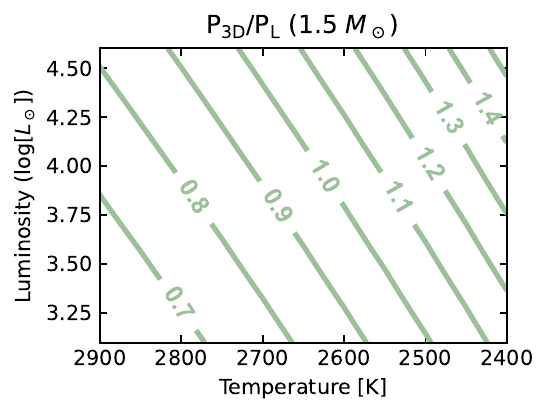} \\
   \includegraphics[width=0.9\hsize]{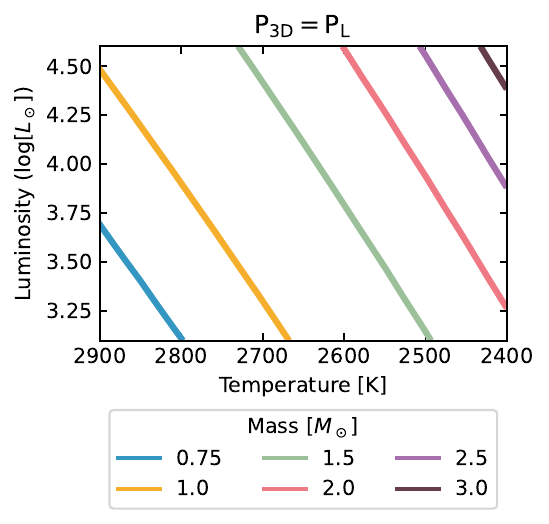}
      \caption{Ratio of pulsation periods, $P_\mathrm{3D}/P_\mathrm{L}$, as a function of stellar parameters. \textit{Top}: Period ratio for a stellar mass of 1.5 $\mathrm{M_\odot}$. \textit{Bottom}: Location of the one-to-one relation ($P_\mathrm{3D}=P_\mathrm{L}$) for different stellar masses.}
         \label{fig:period_ratio}
   \end{figure}

\subsection{Model parameters}

The simulations start with hydrostatic, dust-free atmospheric structures, defined by the stellar parameters current mass $M_\star$, luminosity $L_\star$ and effective temperature $T_\star$ (assuming elemental abundances as specified in Paper I). 
The corresponding stellar radius is given by $R_\star = (L_\star/4\pi \sigma T_\star^{4})^{1/2}$.
The model parameters explored in this paper are based on a subsample of the grid presented in Paper~I. We included the full range of current stellar masses ($M_\star=0.75-3 M_\odot$) but considered narrower ranges in effective temperature ($T_{\star}=2400-2900$ K) and luminosity (range depending on stellar mass and sampled with a step size of $\Delta \log(L_\star)=0.15$). 
Figure~\ref{fig:grid} illustrates the stellar parameter range of the subset of models (coloured circles) selected from the larger grid in Paper~I (grey dots).
The selection was based on targeting models near the wind/no-wind boundary (where the influence of the pulsation period is expected to be largest) and avoiding regions in stellar parameter space with convergence issues 
(see Fig.~6 of Paper~I and the discussion therein). The piston velocity amplitude $\Delta u_\mathrm{p}$ is
 set to 2, 3, and 4~$\mathrm{km,s^{-1}}$ for all combinations of stellar parameters, and the luminosity scaling factor is fixed at $f_\mathrm{L}=2$, following the setup adopted in Paper~I.
The growth of dust grains starts from tiny seed particles, with an assumed abundance that is defined as the ratio of the seed-particle number density, $n_{\rm d}$, to the hydrogen number density, $n_{\rm H}$. Throughout this work, a fixed value of 
$n_\mathrm{d}/n_\mathrm{H} = 3\times10^{-15}$ was adopted, as this value was found to best reproduce observed wind properties (see Paper~I).
All models from Paper~I used in the present study were recomputed to ensure a consistent treatment throughout the analysis. 
A more detailed description of the modelling approach and dust treatment is provided in Paper~I and references therein.

  \begin{figure}[t!]
   \centering
   \includegraphics[width=\hsize]{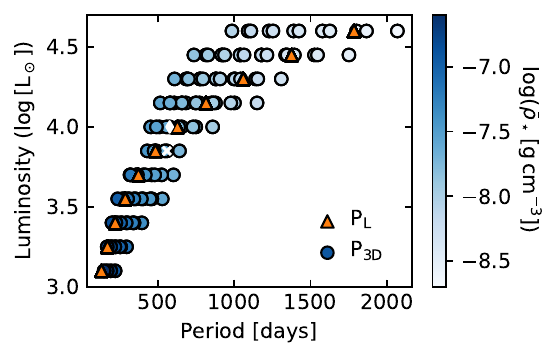}
      \caption{Luminosity plotted against pulsation period, illustrating the distribution of models (coloured circles in Fig.~\ref{fig:grid}) based on the different period relations: the P--L relation \citep[orange triangles;][]{whitelock2009a} and the period--mean density relation \citep[circles coloured by mean density;][]{ahmad2023}. The white symbols identify the two $P_{\rm 3D}$ models selected for the detailed analysis presented in Sect.~\ref{sect:result_dyn}.}
         \label{fig:period_lum}
   \end{figure}

\subsection{Pulsation periods}
\label{sect:method_puls}

The pulsation period of an AGB star can be related to its luminosity through an empirical P--L relation. Such relations have been extensively studied, and several formulations are available in the literature, primarily based on observations of long-period variables in nearby galaxies (e.g. \citealt{feast1989,whitelock2009a}). So far, pulsation periods in DARWIN models have mostly been set according to such observationally grounded P--L relations, avoiding the introduction of period as an extra free parameter. In Paper~I, the relation from \citet{whitelock2009a} was adopted and is therefore used as the baseline for our comparison. The relation is given by

\begin{equation}
    M_\mathrm{bol}=-4.271-3.31(\mathrm{log}P-2.5),
\end{equation}

\noindent with $P$ in days and using a solar $M_{\mathrm{bol}} = 4.74 \, {\mathrm{mag}}$.

More recently, \citet{ahmad2023} used global 3D RHD models of AGB stars, computed with the CO5BOLD code, to investigate how self-excited pulsations arise naturally from internal convective dynamics. The CO5BOLD code numerically solves the coupled, non-linear equations of compressible hydrodynamics and non-local radiative energy transport in a global `star-in-a-box’ configuration, where the entire star is contained within a 3D cubical computational domain. This approach allows large-scale convective flows and acoustic pulsation modes to develop self-consistently through the non-linear interplay between the stellar interior and atmosphere. 
These simulations provide a period–mean density relation that links the pulsation period directly to the internal stellar structure, with the stellar radius defined as the temporally averaged radial position of the innermost local minimum in the entropy profile (see \citealt{ahmad2023} for details).

The significance of this period--mean density relation is that it provides a physically motivated alternative to purely empirical prescriptions. The relation is not a simple power-law fit to observed periods, but is instead anchored in global 3D RHD models in which convection and pulsation are treated as part of the same dynamical system. This is particularly useful for dynamical atmosphere and wind models, where pulsations are often introduced through a prescribed piston boundary condition, requiring the period, phase, and amplitude to be specified externally (see Sect.~\ref{sect:method}). A period--mean density relation resulting from 3D CO5BOLD models links the pulsation period directly to the global stellar structure, through $M_\star$ and $R_\star$, rather than relying on a purely empirical period prescription. 
We adopted the relation for the fundamental radial mode presented in Eq. (5) of \citet{ahmad2023}:

\begin{equation}
\begin{split}
    \log(P_{\mathrm{puls}}) = -0.5625 \log\left(\bar{\rho}_\star/\bar{\rho}_\odot\right)
    -1.6275,
\end{split}
\end{equation}

\noindent with $P_{\mathrm{puls}}$ given in days and the mean density of the star expressed in solar units, such that $\bar{\rho}_\star/\bar{\rho}_\odot=M_\star/R_\star^3$ where $M_\star$ and $R_\star$ are given in units of $M_\odot$ and $R_\odot$, respectively.

In the following, the two prescriptions are denoted with $P_\mathrm{L}$ and $P_{\mathrm{3D}}$ for \citet{whitelock2009a} and \citet{ahmad2023}, respectively. Figure~\ref{fig:period_ratio} compares the two period prescriptions across a range of stellar parameters. At a given luminosity and fixed current mass, the ratio $P_{\rm 3D}/P_{\rm L}$ increases with decreasing effective temperature (the stellar radius increases and the mean stellar density decreases). For a star with a current mass of $1.5~M_\odot$ (top panel), the points where the two periods are equal are found in an effective temperature range of about 2500--2700 K for the luminosity range shown here. At higher temperatures, the ratio falls below unity, indicating that the period–mean density relation gives shorter periods than the P--L relation. The location where the periods are equal also depends on stellar mass, shifting towards lower temperatures for higher masses and towards higher temperatures for lower masses, as shown in the bottom panel.

For the stellar parameters used in our model grid, the corresponding pulsation periods are compared in Fig.~\ref{fig:period_lum}. Using the P--L relation, all models at a given luminosity have the same pulsation period, independent of other stellar parameters. In contrast, the period–mean density relation gives a broader range of periods at a fixed luminosity, reflecting its dependence on mass and radius. 
The relative behaviour of the two prescriptions varies systematically across the parameter space, producing the shift from predominantly $P_{\rm 3D}>P_{\rm L}$ at low luminosities to predominantly $P_{\rm 3D}<P_{\rm L}$ at high luminosities, for combinations of stellar parameters used in this study.
The transition between these regimes occurs around $\log L_\star~[L_\odot] \approx 4$.

Because the velocity amplitude at the inner boundary, $\Delta u_\mathrm{p}$, is kept fixed while the pulsation period is varied, changes in the pulsation period translate directly into changes in the radial amplitude (see Eq.~\ref{eq:R_in}). For the range of models considered here, the relative period difference, $(P_\mathrm{3D}-P_\mathrm{L})/P_\mathrm{L}$, varies between $-45\%$ and $+84\%$. Despite these large differences in period, the resulting difference in the maximum (or minimum) value of $R_{\rm in}(\rm t)$ remains modest: the corresponding difference, $(R_{\mathrm{max}}(P_{\mathrm{3D}})-R_{\rm{max}}(P_\mathrm{L}))/R_0$, lies between $-8\%$ and $+5\%$, where $R_0$ denotes the mean stellar radius and $R_{\rm max}$ denotes the maximum of $R_{\rm in}(\rm t)$. This corresponds to a sensitivity of approximately $0.03$–$0.18\%$ in $R_{\rm max}$ per $1\%$ change in pulsation period. 
The difference translates into only minor changes in the thermal conditions in the dust formation region.

\section{Results}
\label{sect:result}

In this section we compare the resulting wind properties of the two model grids based on different period prescriptions. We start by discussing some representative examples of time-dependent behaviour in detail, to understand the influence of the pulsation period on dust condensation and wind acceleration. Then we compare results for model pairs in the entire grids. A table listing the resulting wind and dust properties of the DARWIN models presented in this paper, together with the corresponding input parameters, is available at the CDS. The data format is described in Table~\ref{tab:app}.

\subsection{Atmospheric structures and time-dependant dynamics}
\label{sect:result_dyn}

\begin{figure}[t!]
   \centering
    \includegraphics[width=0.44\textwidth]{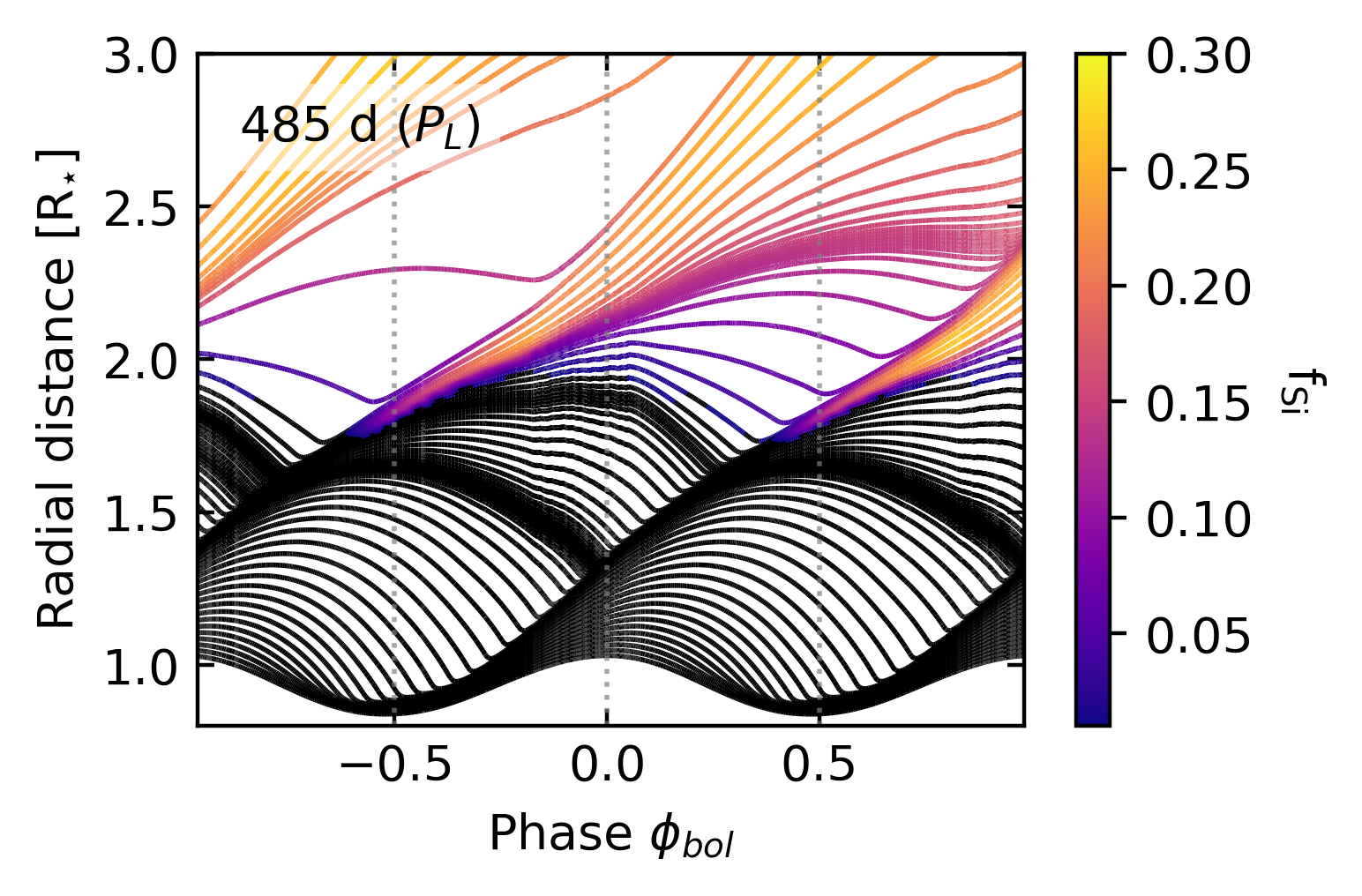} \\
    \includegraphics[width=0.44\textwidth]{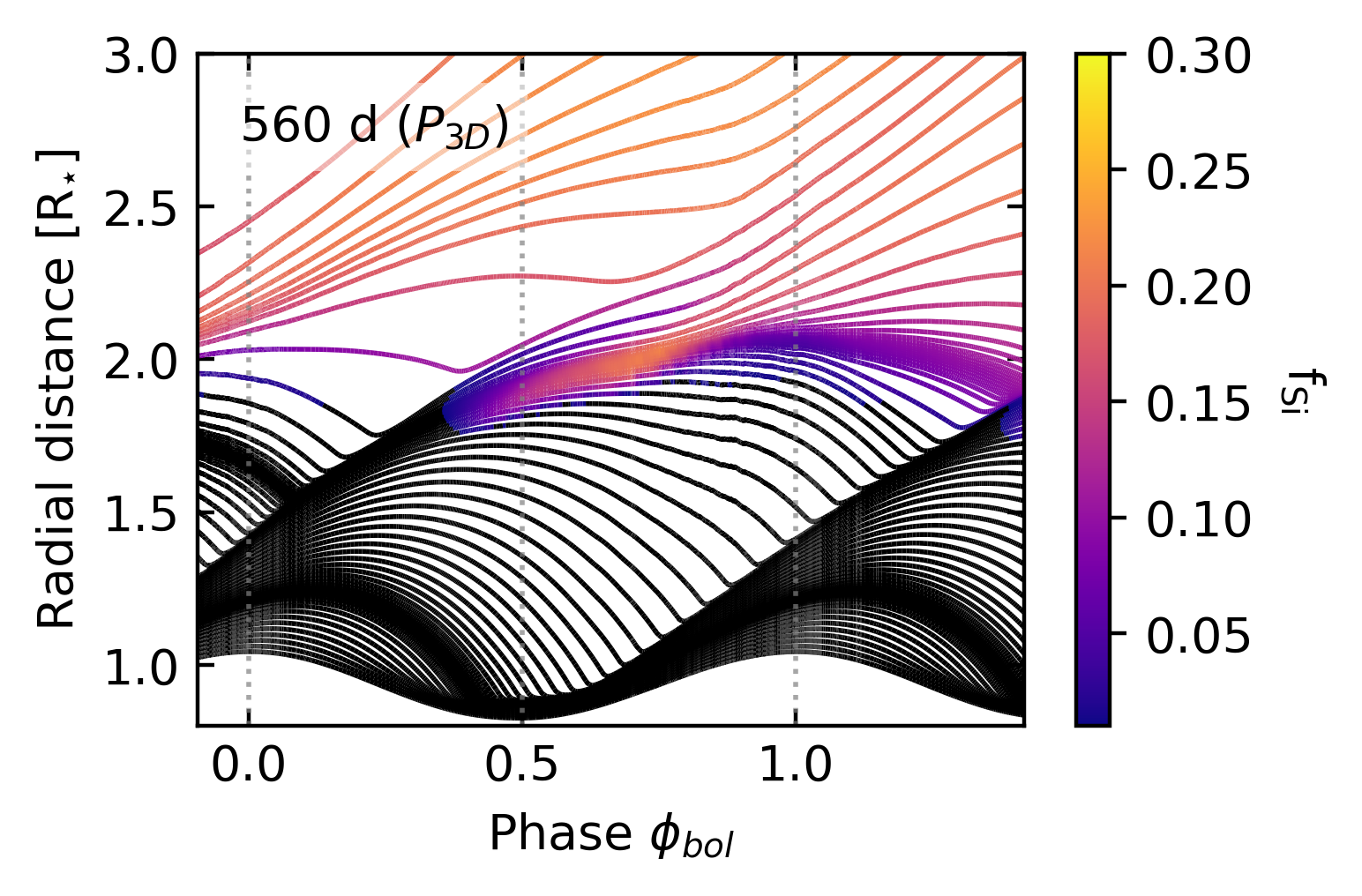}
      \caption[]{Position of selected mass shells\footnotemark \ as a function of time (bolometric phase). Colour represents the condensation fraction of Si (indicating the presence of dust). Anything below a fraction of 0.01 is coloured black. The two plots show models with identical stellar parameters ($7000~L_\odot$, 2500~K, $1.5~M_\odot$, and $\Delta u_\mathrm{p}=4~\mathrm{km\,s^{-1}}$) but different pulsation periods. \textit{Top}: 485 days ($P_\mathrm{L}$). \textit{Bottom}: 560 days ($P_{\mathrm{3D}}$). }
         \label{fig:track1}
   \end{figure}

Figure~\ref{fig:track1} illustrates the dynamics of two models with identical stellar parameters but different pulsation periods ($P_{\rm 3D}$ model marked as a cross in Fig.~\ref{fig:period_lum}). Each curve traces the temporal evolution of a Lagrangian mass shell. The plots show the dust-free inner region dominated by pulsations, the location of dust condensation, and the subsequent outward acceleration of dust by radiation pressure. Shock waves, where infalling material collides with outflowing gas, appear as prominent features.

A comparison of the two models shows that the timing of dust formation within the pulsation cycle strongly influences wind acceleration.
In both cases, new dust forms when overall temperatures are low, around the minimum luminosity phase ($\Phi_{\rm bol}=0.5$). 
For the model with the shorter period ($P_\mathrm{L}$ in this case), a shock wave propagates through the dust formation region, increasing the density of the outward moving material. This enhanced density promotes more efficient dust formation, and the increasing luminosity subsequently accelerates the newly formed dust outwards. 
The most dust-rich layers move away from the star, thereby contributing to the formation of a stellar wind. 
Other material falls back towards the star ballistically, but encounters a new shock wave during the subsequent pulsation cycle, which leads to the next phase of mass loss. For the model with the longer period ($P_\mathrm{3D}$), dust is also formed around minimum light, but somewhat after the passage of the shock wave, leading to less efficient condensation and radiative acceleration. Part of this material is eventually accelerated outwards as the luminosity increases, while a fraction evaporates, and the dust-free gas falls back towards the star.

To quantify the impact on wind properties, we computed time-averaged mass-loss rates, $\dot{M}$, and terminal wind velocities, $u_\infty$, typically over at least 100 pulsation periods. For the two models shown in Fig.~\ref{fig:track1}, the resulting wind properties are $\dot{M}=2.7\times10^{-6}~M_\odot\,\mathrm{yr^{-1}}$ and $v_\infty=16.1~\mathrm{km\,s^{-1}}$ for the model with the shorter period ($P_\mathrm{L}$), and $\dot{M}=1.9\times10^{-6}~M_\odot\,\mathrm{yr^{-1}}$ and $v_\infty=11.3~\mathrm{km\,s^{-1}}$ for the model with the longer period ($P_{\mathrm{3D}}$). 
The model with more favourable dust condensation timing produces both a higher mass-loss rate and a higher wind velocity.

\begin{figure}[t!]
    \centering
    \includegraphics[width=0.44\textwidth]{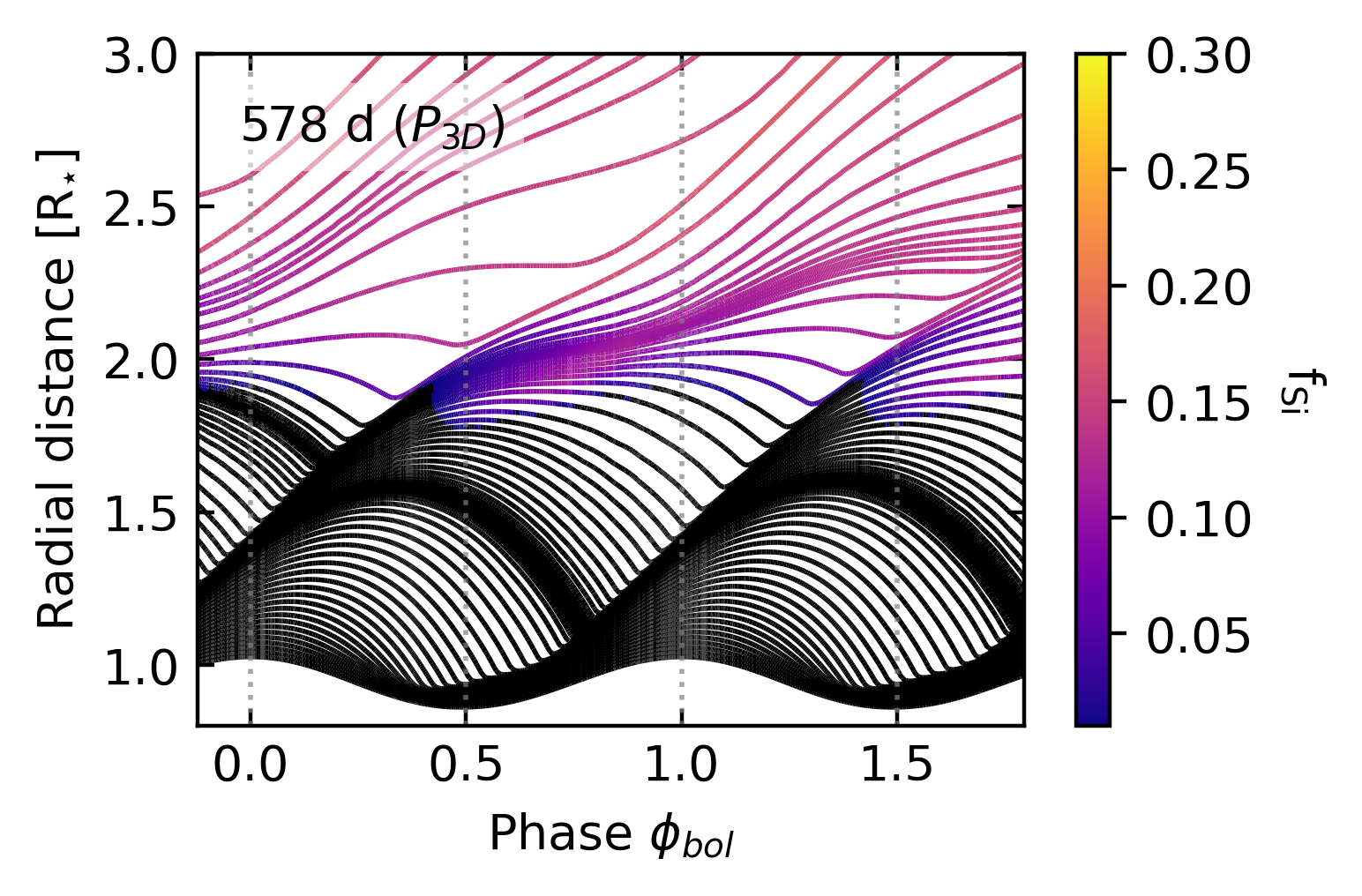} \\
    \includegraphics[width=0.44\textwidth]{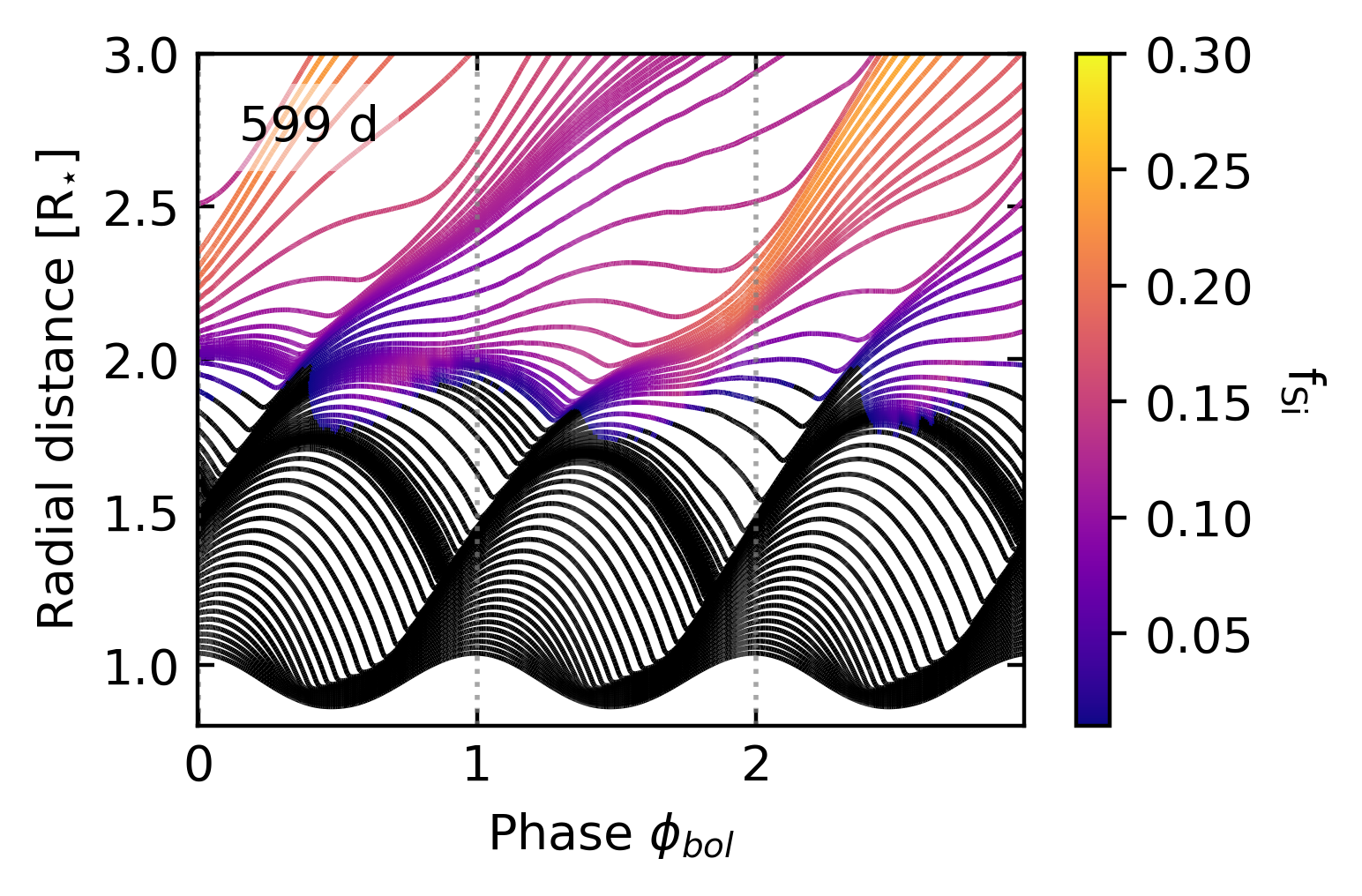} \\
    \includegraphics[width=0.44\textwidth]{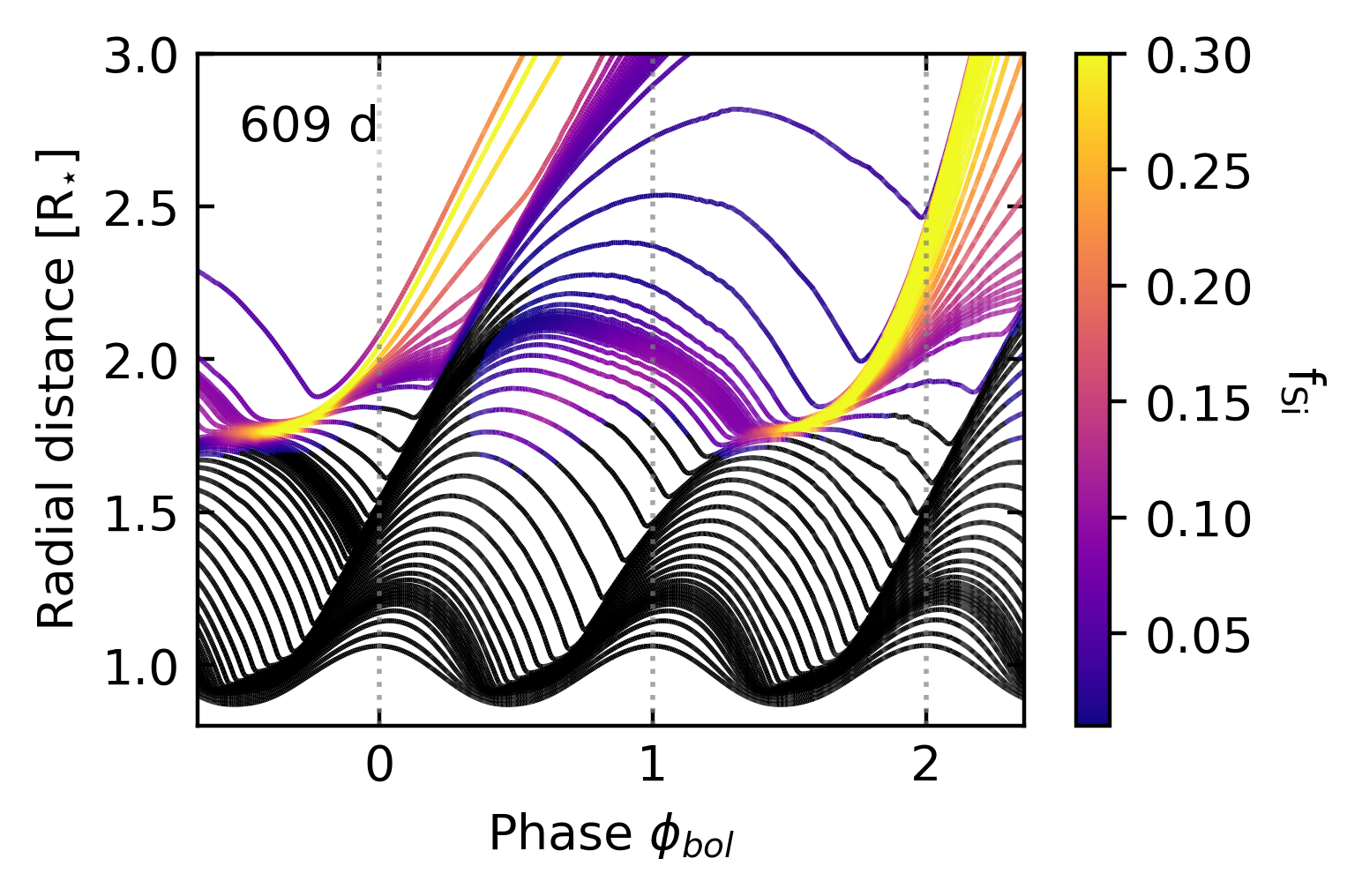} \\
    \includegraphics[width=0.44\textwidth]{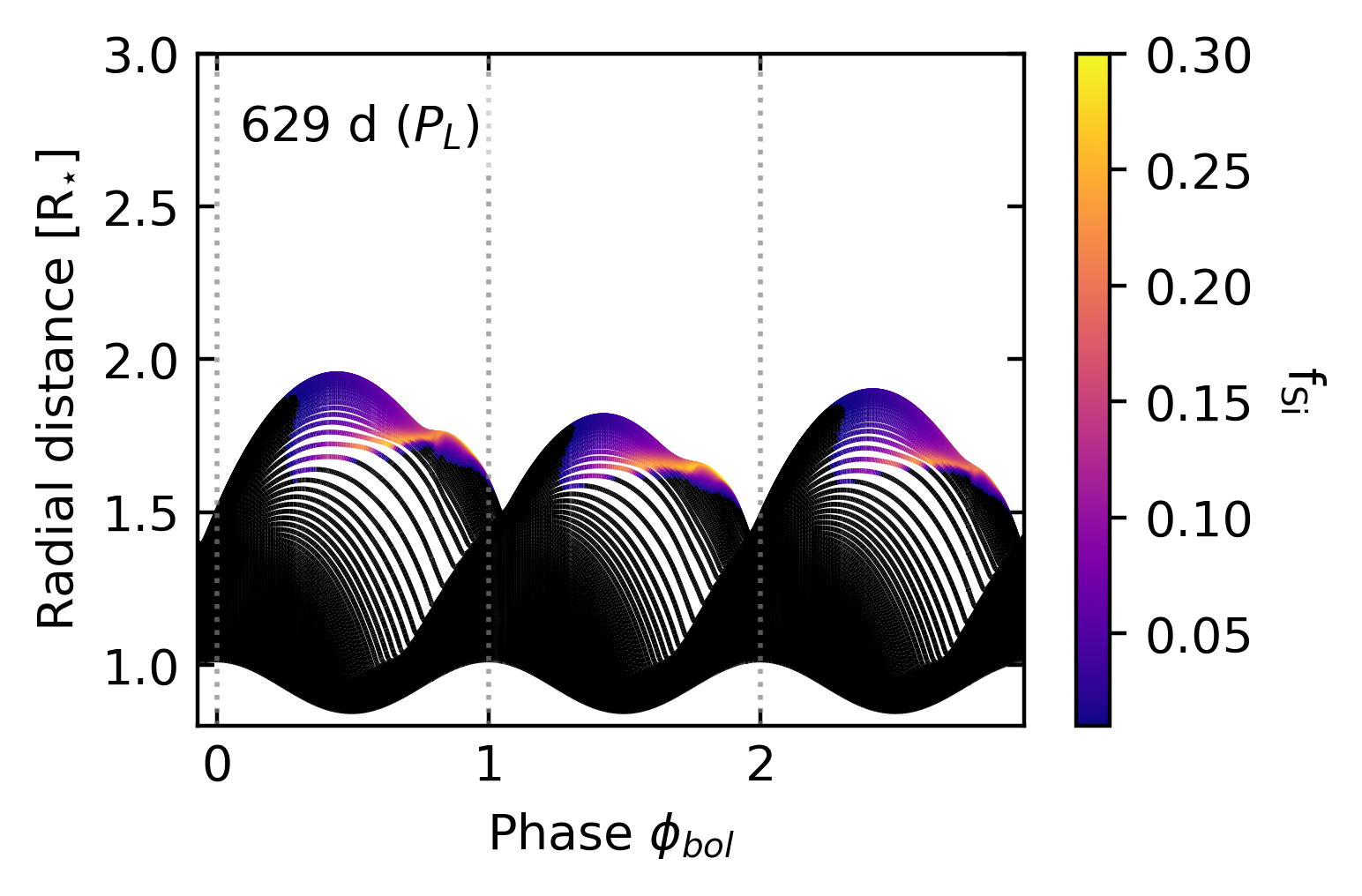}
    \caption{Mass shell plots for the models in Table~\ref{tab:periodseries} ($10\,000~L_\odot$, $2700$~K, $1.5~M_\odot$, and $\Delta u_\mathrm{p}=3~\mathrm{km\,s^{-1}}$) with different pulsation periods, from 578 days ($P_\mathrm{3D}$) to 629 days ($P_\mathrm{L}$). See Fig.~\ref{fig:track1} for details. }
    \label{fig:track2}
\end{figure}

In Fig.~\ref{fig:track2} we present 
a model pair in which only the $P_{\rm 3D}$ model produces a wind ($P_{\rm 3D}$ model marked as a diamond in Fig.~\ref{fig:period_lum}).
To investigate the transition between wind-producing and non-wind models for this particular case, we computed a series of models with intermediate periods and fixed stellar parameters. The resulting wind properties along this sequence are summarised in Table~\ref{tab:periodseries}. Since the models with periods of 609 and 614 days exhibit a similar dynamical behaviour, only one of them is shown in Fig.~\ref{fig:track2}.

\footnotetext{For clarity, we note that the selection of trajectories does not correspond to physical quantities such as gas density or other local model properties. Instead, the tracks originate from grid points at a given snapshot that are used as starting positions to follow the motion of the gas. Their spatial distribution reflects the behaviour of the adaptive grid, which concentrates grid points in regions with steep gradients. For the purpose of this figure, the distribution of trajectories can therefore be regarded as effectively random. We also note that the temporal resolution of the tracks is limited by the time steps used in the model computation, which can occasionally be relatively large.}

\begin{table}[t]
    \caption{Average wind properties for models with fixed stellar parameters ($10\,000~L_\odot$, $2700$~K, $1.5~M_\odot$, and $\Delta u_\mathrm{p}=3~\mathrm{km\,s^{-1}}$) but different periods.}
    \centering
    \begin{tabular}{c|c|c|c|c} \hline\hline
        $P$ & $\dot{M}$ & $u_\infty$ & $f_{\mathrm{Si}}$ & $a_\mathrm{gr}$ \\
        $\mathrm{[d]}$ & [$M_\odot\,\mathrm{yr^{-1}}$] & [$\mathrm{km\,s^{-1}}$] & & [$\mathrm{\mu m}$] \\ \hline
        578 & $1.9\times10^{-6}$ & 12.5 & 0.18 & 0.33 \\
        599 & $1.2\times10^{-6}$ & 16.9 & 0.23 & 0.36 \\
        609 & $9.3\times10^{-7}$ & 17.5 & 0.24 & 0.36 \\ 
        614 & $8.0\times10^{-7}$ & 16.0 & 0.23 & 0.36 \\
        619 & -- & -- & -- & -- \\
        629 & -- & -- & -- & -- \\ \hline
    \end{tabular}
    \tablefoot{The columns show pulsation period $P$, mass-loss rate $\dot{M}$, wind velocity at the outer boundary $u_\infty$, condensation fraction of silicon $f_{\rm Si}$ and grain radius $a_{\rm gr}$. Blank entries indicate models that do not sustain a wind.}
    \label{tab:periodseries}
\end{table}

In the upper panel (period 578 days), the behaviour resembles that of the second model in Fig.~\ref{fig:track1}: part of the newly formed dust evaporates, while another fraction is accelerated outwards and contributes to a wind. For the model with a period of 599 days, we find similar behaviour but with a higher degree of condensation. This can be attributed to the longer period, which results in an extended phase of lower temperatures, allowing the dust to grow to sizes optimal for efficient radiative acceleration (see e.g. Fig.~1 in \citealt{hofner2008}). Combined with a longer phase of high luminosity, i.e. a longer interval of efficient radiative acceleration, this leads to a higher wind velocity (see Table~\ref{tab:periodseries}).

We also find indications that the degree of condensation is higher every second pulsation cycle. This effect becomes more pronounced in the model with a period of 609 days, where every second cycle leads to a substantial increase in condensation degree and a corresponding rapid acceleration around luminosity maximum. 
Over longer time series, this model (as well as the model with a 614-day period) exhibits large temporal variations in wind properties compared to the first model, with extended phases of reduced condensation efficiency followed by periods of alternating high and low condensation from cycle to cycle. This means that the time-averaged wind properties reflect a larger intrinsic variability in these models compared to the 578-day model.

In this series of models, increasing the period initially leads to a higher degree of condensation and higher wind velocities. For longer periods, however, the degree of condensation and wind velocity remain approximately constant, while the mass-loss rate decreases from model to model. For even longer periods, the behaviour changes: the newly formed dust evaporates before radiation pressure or shock waves can accelerate the material outwards, preventing the formation of a sustained wind.

   \begin{figure}[t!]
   \centering
    \includegraphics[width=0.92\hsize]{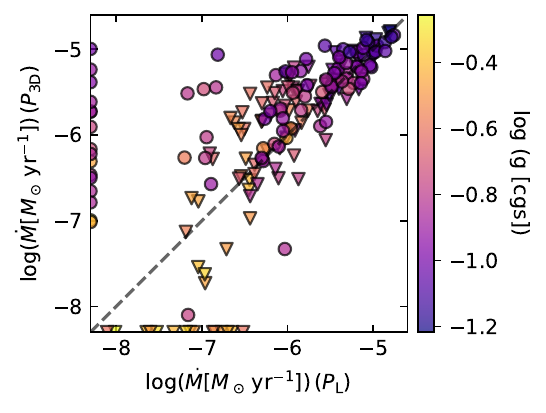} \\
    \includegraphics[width=0.83\hsize]{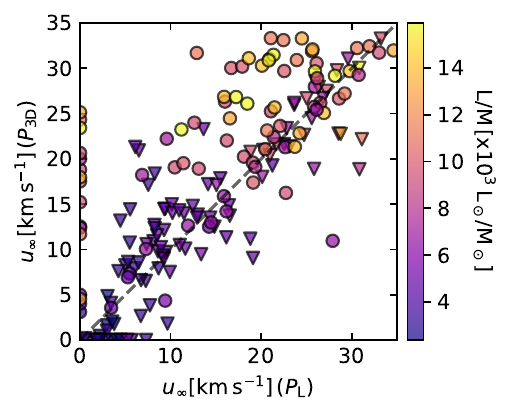} \\ 
    \includegraphics[width=0.92\hsize]{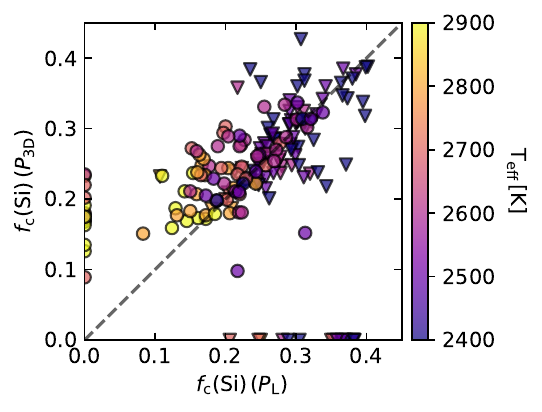}
      \caption{Comparison of time-averaged wind properties. Values of the $P_\mathrm{L}$ model are on the x-axis and the $P_\mathrm{3D}$ model on the y-axis, with the symbols indicating which of the two models has the shortest period: $P_\mathrm{L}$ (triangle) or $P_\mathrm{3D}$ (circle). \textit{Top}: Mass-loss rates.  \textit{Middle}: Wind velocities. \textit{Bottom}: Degree of condensed Si. Model pairs in which only one model produces a wind are plotted along the axes corresponding to the wind-producing cases. In particular, in the top panel, no-wind cases ($\dot{M}=0$) cannot be represented on the logarithmic scale and are therefore displayed on the corresponding axes.}
         \label{fig:wind_pdiff}
   \end{figure}

\begin{figure*}[htb!]
    \centering
    \includegraphics[width=0.48\linewidth]{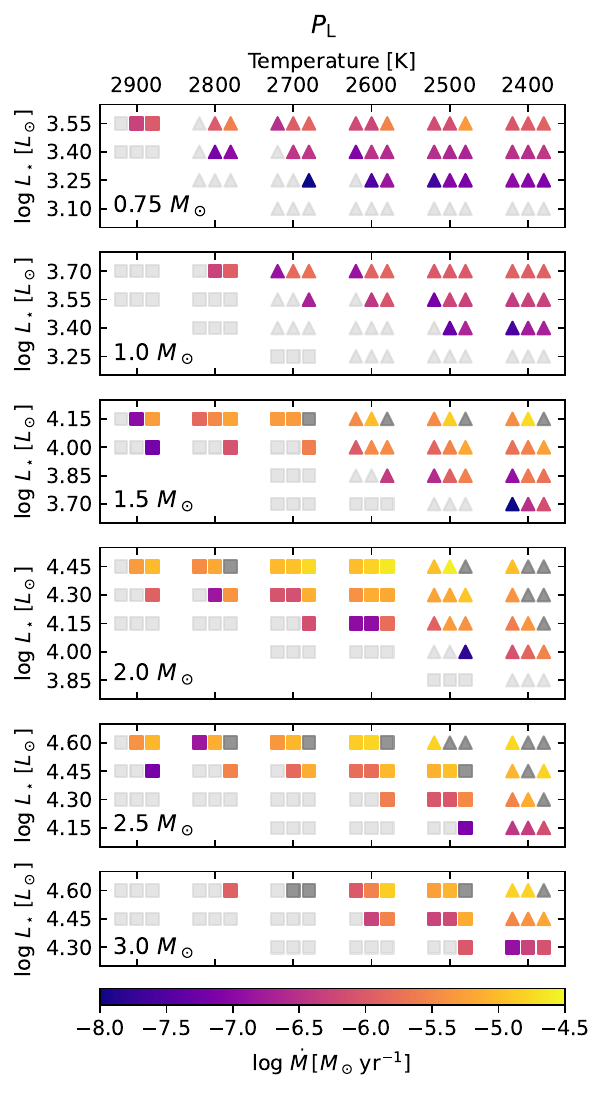}
    \includegraphics[width=0.48\linewidth]{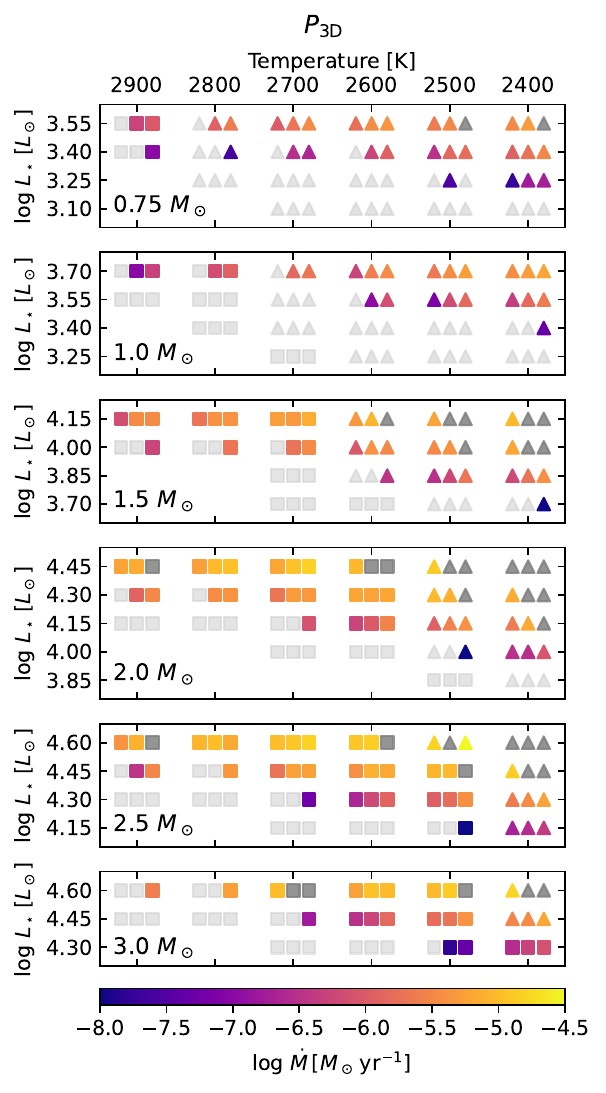}
    \caption{Schematic overview of the models in the grid as a function of model parameters, coloured by time-averaged mass-loss rates. The panels show different current stellar masses. The three adjacent symbols at each temperature represent the piston velocities $\Delta u_\mathrm{p}=2$, 3, and 4 $\mathrm{km\,s^{-1}}$ from left to right. Light grey symbols represent models with no wind, and dark grey symbols represent models that develop a wind but fail to converge for numerical reasons. The symbol shows which period is shorter for each parameter combination: $P_\mathrm{L}<P_\mathrm{3D}$ (triangles) or $P_\mathrm{3D}<P_\mathrm{L}$ (squares). \textit{Left}: All models with period $P_\mathrm{L}$. \textit{Right}: All models with period $P_\mathrm{3D}$.  }
    \label{fig:wm1}
\end{figure*}

\subsection{Comparison of average wind properties}

Figure~\ref{fig:wind_pdiff} presents time-averaged mass-loss rates and wind velocities for all computed model pairs that differ only in pulsation period. To indicate which model in each pair has the shorter period, cases with $P_\mathrm{L}<P_\mathrm{3D}$ are shown as triangles, while cases with $P_\mathrm{3D}<P_\mathrm{L}$ are shown as circles. 
Models aligned along the axes correspond to cases in which a wind develops for only one of the two periods considered. In all such cases, it is consistently the shorter period that produces a wind (as indicated by the symbols).
This behaviour suggests the existence of an upper limit in the pulsation period beyond which the timing of dust formation becomes unfavourable: dust forms too late in the pulsation cycle for efficient radiative acceleration, falls back towards the star, and evaporates before contributing effectively to wind driving.

For the mass-loss rates (top panel), the models are colour-coded by surface gravity ($\log g$), revealing a clear correlation. Low surface gravity plays a key role in enabling dust-driven mass loss: weaker gravity allows pulsation-generated shocks to levitate more gas to distances where dust can condense, after which radiation pressure on the dust accelerates the material beyond the escape velocity. While the models show a substantial spread overall, they cluster more closely around the one-to-one relation at higher mass-loss rates. This indicates that the influence of the pulsation period diminishes with lower surface gravity.

The wind velocities (middle panel of Fig.~\ref{fig:wind_pdiff}) exhibit a more complex behaviour, depending on which of the two periods is shorter. Models where $P_\mathrm{L}<P_\mathrm{3D}$ (triangles), primarily corresponding to lower-mass and lower-luminosity cases (see Figs.~\ref{fig:period_ratio} and \ref{fig:period_lum}), show a wide spread without a clear trend. In contrast, models with $P_\mathrm{3D}<P_\mathrm{L}$ (circles), typically associated with higher masses and luminosities, tend either to lie close to the one-to-one relation or to display higher wind velocities for the shorter period ($P_\mathrm{3D}$). 
The colour-coding by $L_\star/M_\star$, which is related to the ratio of radiative to gravitational acceleration ($a_{\rm rad}/a_{\rm grav}$), suggests that for models with higher $L_\star/M_\star$, and therefore more favourable conditions for radiative driving, 
a shorter pulsation period tends to enhance the wind velocity. 

The condensation fractions (bottom panel) show no clear dependence on the adopted pulsation period. Instead, they correlate strongly with effective temperature: lower temperatures result in higher condensation fractions, reflecting more favourable conditions for dust formation. 

The effects discussed above highlight the complexity of the wind-driving process. There is no simple trend with the adopted prescription of pulsation period; instead, the resulting wind properties depend on a combination of timing, density structure, and dust formation efficiency.

\subsection{Mass-loss rate versus stellar parameters}
\label{sect:mlossvspar}

Figure~\ref{fig:wm1} provides an overview of how the mass-loss rate depends on stellar parameters and pulsation period. Each combination of stellar mass, luminosity, effective temperature, and piston velocity is represented by a symbol, colour-coded by the time-averaged mass-loss rate. The different pulsation periods are shown in separate panels (left: $P_\mathrm{L}$; right: $P_\mathrm{3D}$). The figure highlights the location of the wind/no-wind boundary in parameter space and how it shifts with the choice of period.
A clear result is that the position of this boundary depends on the pulsation period. As noted in the previous section, shorter periods generally favour wind formation. This is reflected in the figure as an extension of the wind-producing models into regions of parameter space that otherwise correspond to no-wind cases.

This effect depends on stellar mass. For the $0.75$ and $1.0\,M_\odot$ models, where the P--L relation typically gives shorter periods than the period--mean density relation, the wind regime extends to higher luminosities and effective temperatures in the $P_\mathrm{L}$ case. At higher masses, where the $P_\mathrm{3D}$ relation instead results in shorter periods, the trend is reversed: more wind-producing models are found when adopting the period--mean density relation. In other words, changing the period prescription shifts the wind/no-wind boundary in different directions across the parameter space, depending on the stellar mass range.

As discussed in Paper~I (Sect.~4.1), the DARWIN models are currently biased against cases with very high mass-loss rates, where dust formation and wind acceleration are very abrupt, leading to convergence issues. This is also evident in Fig.~\ref{fig:wm1}, where non-converged models (dark grey) are located in the upper-right region of the parameter space at higher masses, typically adjacent to models with the highest mass-loss rates.

\begin{figure}[t!]
    \centering
    \includegraphics[width=0.95\hsize]{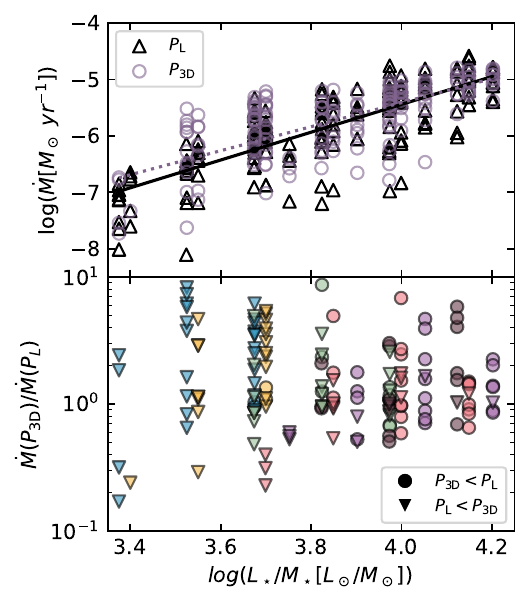}
    \caption{Mass-loss rate as a function of $L_\star/M_\star$. \textit{Top}: All wind-forming models. The solid and dotted lines represent the linear fit to the $P_\mathrm{L}$ and $P_\mathrm{3D}$ models, respectively. \textit{Bottom}: Ratio of mass-loss rates for model pairs, $\dot{M}(P_{\mathrm{3D}})/\dot{M}(P_\mathrm{L})$, colour-coded by mass; see Figs.~\ref{fig:grid} and \ref{fig:period_ratio}.}
    \label{fig:loverm}
\end{figure}

The dependence of mass loss on stellar parameters was examined in detail in Paper~I, where luminosity was identified as the single parameter showing the strongest correlation: higher luminosities correspond to higher mass-loss rates. This trend is particularly evident when considering each stellar mass separately. 
We find that differences in pulsation period can lead to significant deviations in mass-loss rates; however, these differences are distributed across the full luminosity range and do not alter the overall correlation between mass-loss rate and luminosity. In Paper~I, no clear correlation (of mass-loss rate) with effective temperature was identified, and this remains the case here. Although model pairs exhibit differences in mass-loss rate across the full temperature range, no systematic trend is present.
In Fig.~\ref{fig:loverm} the mass-loss rate is shown as a function of $L_\star/M_\star$ for all wind-producing models (upper panel), together with 
the ratio of mass-loss rates for model pairs (lower panel). As discussed in Paper I, the mass-loss rate increases systematically with $L_\star/M_\star$. Mass-loss rate ratios of models with different pulsation periods reach up to one dex, but show no systematic dependence on $L_\star/M_\star$, leading to a similar trend for the two sets of models, as seen in the upper panel.

\begin{figure}[t!]
   \centering
    \includegraphics[width=\hsize]{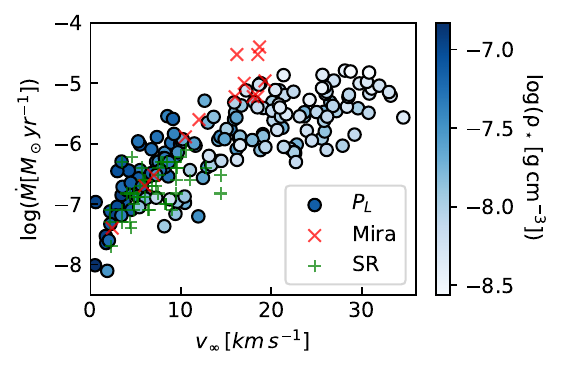} \\
    \includegraphics[width=\hsize]{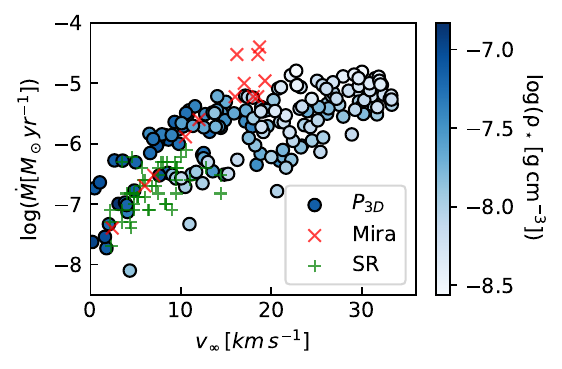}
      \caption{Mass-loss rates vs wind velocities for models that form a stellar wind and the corresponding properties from observed M-type AGB stars, derived from CO lines \citep{olofsson2002, gonzalezdelgado2003}. Models are colour-coded according to mean density ($\log (\bar{\rho}_\star)$). \textit{Top}: Models based on the P--L relation. \textit{Bottom}: Models based on the period--mean density relation.}
         \label{fig:mdot_uinf}
   \end{figure}

\section{Comparison with observations and discussion}
\label{sect:disc}

\subsection{Comparison to observations}

Earlier DARWIN models have shown good agreement with observations in reproducing combinations of mass-loss rates and wind velocities. When comparing the model grid with observations, it is important to keep in mind that  
the grid samples the parameter space uniformly. Consequently, the relative numbers of models with different stellar parameter combinations do not reflect the expected frequency of stars with these properties in an actual AGB population or observed sample.
In Fig.~\ref{fig:mdot_uinf} we show the resulting mass-loss rates and wind velocities for all wind-producing models together with observational values adapted from \citet{olofsson2002} and \citet{gonzalezdelgado2003}. The model results are colour-coded by mean stellar density, and the observations are classified as Mira variables and semi-regular variables (SRVs). As discussed in the previous section, individual models can be strongly sensitive to the adopted pulsation period, an effect that is also reflected in the differences between the two panels of this figure. Nevertheless, the overall distribution of wind properties is broadly consistent with the observed values for both types of models, although the highest observed mass-loss rates are not reached by the current models due to numerical limitations (convergence issues for combinations of high luminosities and low effective temperatures), as mentioned in the previous section. 

A clear correlation with mean density is apparent, with lower mean density models generally producing faster winds with higher mass-loss rates. The most pronounced differences between the two period prescriptions occur at the highest mean densities ($\log(\bar{\rho}_\star \, [\mathrm{g\,cm}^{-3}]) \approx -7$). In the $P_\mathrm{L}$ case (top panel), these models cluster below 10 $\mathrm{km\,s^{-1}}$, whereas in the period--mean density case (bottom panel) they are more widely distributed, reaching wind velocities up to 15 $\mathrm{km\,s^{-1}}$ and slightly higher mass-loss rates.

\begin{figure}[t!]
   \centering
    \includegraphics[width=\hsize]{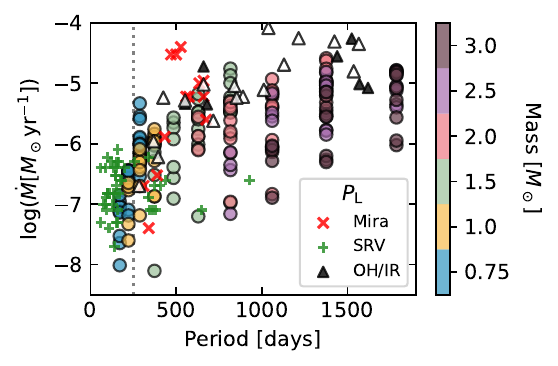} \\
    \includegraphics[width=\hsize]{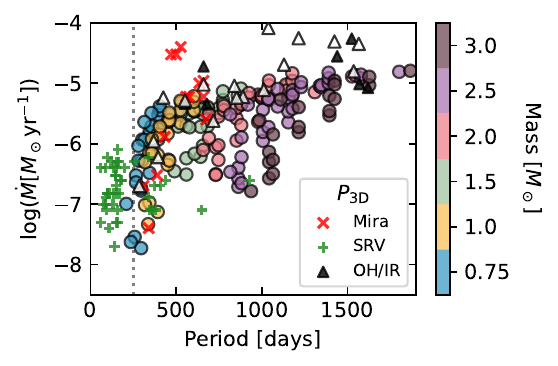}
      \caption{Mass-loss rates vs period for models that form a stellar wind and the corresponding properties from observed M-type AGB stars: Miras and SRVs  \citep{olofsson2002,gonzalezdelgado2003} and OH/IR stars from \cite{debeck2010} (filled triangles) and \citet{he2001} (open triangles). Models are colour-coded according to current mass. \textit{Top}: Models based on the P--L relation. \textit{Bottom}: Models based on the period--mean density relation.}
         \label{fig:mdot_p}
   \end{figure}

Figure~\ref{fig:mdot_p} shows the mass-loss rate as a function of pulsation period for all wind-producing models, together with observational samples from \citet{olofsson2002}, \citet{gonzalezdelgado2003}, \citet{he2001}, and \citet{debeck2010}. The mass-loss rates are derived from CO-line emission, except for the sample from \citet{he2001}, where dust emission, assuming a dust-to-gas mass ratio of $10^{-3}$, is used. In the observational sample, the Miras and SRVs form two sequences separated by the dotted vertical line. As discussed in Paper~I, the stars with shorter periods are most likely first-overtone pulsators, whereas the stars with longer periods pulsate in the fundamental mode.

Nearly all wind-producing models using the period--mean density relation ($P_\mathrm{3D}$) lie to the right of this line, whereas several models based on the P--L relation ($P_\mathrm{L}$) lie to the left. Our present model setup is intended to resemble Mira variables pulsating in the fundamental mode, and adopting the period--mean density relation results in pulsation periods that are more consistent with this assumption. By contrast, with the P--L relation, several wind models have periods closer to the first-overtone sequence.

In \citet{debeck2010}, a linear relation is fitted to the sample for periods below 850 days, and it agrees well with the data of \citet{olofsson2002} and \citet{gonzalezdelgado2003} as well as with both sets of models. For longer periods ($P>850$ days), \citet{debeck2010} find that the mass-loss rate levels off to a constant value of $\log(\dot{M}\,[M_\odot\,\mathrm{yr}^{-1}]) = -4.5$. 
A similar saturation is seen in the DARWIN models, although at lower mass-loss rates than observed. As mentioned before, this behaviour may partly reflect limitations in the parameter space covered by the model grid (convergence issues; see Sect.~\ref{sect:mlossvspar} and Fig.~\ref{fig:wm1}).

\subsection{Pulsation modes}

Although the present DARWIN sub-grid is constructed to represent AGB stars
pulsating predominantly in the fundamental mode, the observational context
against which we compared wind properties encompasses both
Miras and SRVs, with the latter often associated with overtone
pulsations. 
Some overlap between the fundamental and first overtone regimes is expected on physical grounds. Global 3D RHD models computed with CO5BOLD show that, in the low mean-density regime characteristic of luminous, extended AGB 
atmospheres, the dynamical timescale is close to the period of the fundamental acoustic mode \citep{ahmad2023}; this leads to high amplitudes of the fundamental 
mode pulsations. The amplitudes of overtone modes become larger with increasing mean density \citep{ahmad2025}.

In \citet{ahmad2025}, the transition from fundamental to first overtone dominance occurs
around a mean density of $\log(\bar{\rho}_\star \, [\mathrm{gcm}^{-3}]) = -6.8$ corresponding to a luminosity of $\log(L_\star \, [L_\odot]) \approx 3.5$. 
Here, the mean density is defined as $\bar{\rho}_\star = {3M_\star}/{4\pi {R}_\star^3}$, where $M_\star$ is the current stellar mass and ${R}_\star$ is the temporally averaged stellar radius in the 3D RHD model. 
At each snapshot, an instantaneous radius $R_{\star}(t)$ is determined from the radial position of the innermost local minimum in the spherically averaged entropy profile. This radius varies with pulsation phase as the star contracts and expands. The value used in the mean-density calculation is therefore the temporal average of $R_{\star}(t)$, such that the definition accounts for pulsation-induced radius variations. This entropy-minimum criterion was adopted because it provides the most consistent boundary for the stellar radius across the 3D model grid (see \citealt{ahmad2023}), marking the transition between the convective interior and the overlying atmosphere.
The quantity $d\log P/d\log\bar{\rho}_\star$ then denotes the slope of the fitted period–mean density relation in logarithmic space, and quantifies how sensitively the pulsation period $P$ varies with the mean density.
In the low mean-density regime, the fitted slope of
$-0.539 \pm 0.052$, remains consistent with 
the classical mean-density scaling for the radial fundamental mode by \citet{ritter1879}, which corresponds to a slope of $-0.5$. Above 
$\log(\bar{\rho}_\star \, [\mathrm{gcm}^{-3}]) = -6.8$,
the gradient steepens to $-0.621 \pm 0.023$ for the fundamental mode,
indicating a systematic change in the pulsation behaviour as the stellar structure 
becomes denser. Modest changes in the effective pulsation timescale can shift the phase relation between shock propagation, atmospheric levitation, and dust formation, making wind models sensitive to the adopted period prescription. Section \ref{sect:result_dyn} shows how modest changes in the adopted pulsation period can modify the wind properties, including cases where only the shorter period yields a wind. The period prescription affects the wind solution through the timing of the imposed variability relative to atmospheric and dust-formation timescales, and hence through possible resonance effects that can alter the efficiency of dust production and momentum transfer.

For the present grid, the parameter range is not expected to extend far beyond the fundamental-mode dominated regime identified in \citet{ahmad2025}. In the transition region where the first overtone amplitude becomes comparable, multi-periodicity and mode interference can modulate the variability and, potentially, the efficiency of atmospheric levitation and dust production. A more complete assessment of how this transition regime affects mass loss will require a framework that treats mode selection and multi-periodicity explicitly, for example by coupling to pulsation models that predict mode stability and growth rates and by testing how these alter wind initiation near the boundary between wind and no-wind solutions.

\section{Conclusions}
\label{sect:concl}

We have computed two grids of DARWIN models for M-type AGB stars to investigate the effect of different pulsation period prescriptions on the resulting wind properties, comparing models using periods based on the P--L relation with those derived from the period--mean density relation. Applying the period--mean density relation leads to systematic differences in the assumed pulsation periods across the grid, with shorter periods for higher-mass (and higher-luminosity) models and longer periods for lower-mass (and lower-luminosity) models. For the stellar parameter ranges considered here, this prescription corresponds to periods that are more consistent with fundamental-mode Mira variables than those obtained from the P--L relation.

The influence of the pulsation period on wind properties is complex. Differences in assumed periods do not result in simple trends in quantities such as mass-loss rate and wind velocity, particularly near the wind/no-wind boundary. In this region, even modest differences in period can be critical for whether a model develops a wind or not. Well within the wind regime, however, the sensitivity to period is reduced.
Despite these differences, the global trends with stellar parameters remain similar. In particular, the mass-loss rate correlates systematically with $L_\star/M_\star$, indicating that the overall efficiency of radiative driving relative to gravity governs the time-averaged wind properties. The pulsation period mainly regulates the conditions for dust formation and influences the detailed dynamical behaviour of individual models.

Shorter periods are generally more favourable for the onset of a wind and can, in some cases, lead to higher wind velocities, especially for models with lower mean densities. In addition, some models exhibit pronounced cycle-to-cycle or long-term variability, including alternating behaviour between pulsation cycles and extended phases of reduced condensation efficiency, which can significantly affect the time-averaged wind properties. Such behaviour preferentially occurs in models with longer pulsation periods approaching the wind/no-wind boundary.

Both period prescriptions lead to results that are consistent with the combinations of observed mass-loss rates and wind velocities, although the highest observed mass-loss rates are not currently covered by the models due to numerical limitations. The period–mean density relation provides a physically consistent framework for assigning pulsation periods, also taking effects of stellar parameters other than luminosity into account, and therefore enables a more meaningful comparison between DARWIN models and observed Mira variables. Overall, the results highlight the interplay between pulsation, atmospheric structure, and dust formation in shaping AGB star winds. While the pulsation period influences the onset and variability of the outflow, the time-averaged wind properties are primarily determined by the relative strengths of radiative acceleration and gravity.

\section*{Data availability}
Table~\ref{tab:app} is only available in electronic form at the CDS via anonymous ftp to cdsarc.u-strasbg.fr (130.79.128.5) or via http://cdsweb.u-strasbg.fr/cgi-bin/qcat?J/A+A/.

\begin{acknowledgements}
      We thank Sara Bladh for making the original grid of hydrostatic and dynamical models of Paper I available for this study. This work is part of a project that has received funding from the European Research Council (ERC) under the European Union’s Horizon 2020 research and innovation programme (Grant agreement No. 883867, project EXWINGS), and the Swedish Research Council (Vetenskapsrådet, grant number 2019-04059).
      The computations were enabled by resources provided by the National Academic Infrastructure for Supercomputing in Sweden (NAISS) and the Swedish National Infrastructure for Computing (SNIC) at UPPMAX partially funded by the Swedish Research Council through grant agreements no. 2022-06725 and no. 2018-05973.
\end{acknowledgements}


\bibliographystyle{aa}
\bibliography{bibliography}

\begin{appendix}

\onecolumn
\section{Model parameters and results}

\begin{table*}[h!]
    \caption{Input parameters of the models and the resulting averaged wind and dust properties for the two period prescriptions (extract).}
    \centering
    \begin{tabular}{c c c c|c c c c c|c c c c c} \hline \hline
        $M_\star$ & log $L_\star$ & $T_\star$ & $u_\mathrm{p}$ & $P_\mathrm{L}$
        & $\dot{M}$ & $u_\infty$ & $f_\mathrm{Si}$ & $a_\mathrm{gr}$ & $P_\mathrm{3D}$
        & $\dot{M}$ & $u_\infty$ & $f_\mathrm{Si}$ & $a_\mathrm{gr}$\\
        $[M_\odot]$ & $[L_\odot]$ & [K] & $[\mathrm{km\,s^{-1}}]$ & [d]
        & [$M_\odot\,\mathrm{yr^{-1}}$] & [$\mathrm{km\,s^{-1}}$] & & [$\mathrm{\mu m}$] & [d]
        & [$M_\odot\,\mathrm{yr^{-1}}$] & [$\mathrm{km\,s^{-1}}$] & & [$\mathrm{\mu m}$] \\ \hline
        0.75 & 3.25 & 2400 & 2.0 & 171 & 1.09E-07 & 3.8 & 0.40 & 0.43 
        & 296 & 1.86E-08 & 1.8 & 0.39 & 0.43 \\ \hline
    \end{tabular}
    \label{tab:app}
    \tablefoot{The full table is available at the CDS. A portion is shown here for guidance regarding its form and content. The columns list the input stellar parameters (current stellar mass $M_\star$, stellar luminosity $L_\star$, effective temperature $T_\star$, and piston velocity $u_{\rm p}$). These are followed by the resulting wind and dust properties (mass-loss rate $\dot{M}$, wind velocity $u_\infty$, condensation fraction $f_{\rm Si}$ and grain radius $a_{\rm gr}$) for models using the P--L relation $P_{\rm L}$, and the period--mean density relation $P_{\rm 3D}$, respectively. }
\end{table*}

\twocolumn

\end{appendix}
\end{document}